\documentclass[aps,pra,10pt,groupedaddress,showpacs]{aipproc}
\usepackage{amssymb}
\usepackage{amsmath}
\usepackage{graphicx}
\usepackage{array}

\layoutstyle{6x9}

\begin{document}

\title{The necessity of entanglement and the equivalency of Bell's theorem with the second law of thermodynamics}

\author{Ian T. Durham}{
address={Department of Physics, Saint Anselm College, Manchester, NH 03102},
email={idurham@anselm.edu}}
\date{\today}


\begin{abstract}
We demonstrate that both Wigner's form of Bell's inequalities as well as a form of the second law of thermodynamics, as manifest in Carath\'{e}odory's principle, can be derived from the same simple experimental and statistical mechanical assumptions combined with the trivial behavior of integers.  This suggests that Bell's theorem is merely a well-disguised statement of the second law.  It also suggests that entanglement is necessary for quantum theory to be in full accord with the second law and thus builds on the results of Wiesniak, Vedral, and Brukner \cite{Marcin-Wiesniak:2008fv} who showed it was necessary for consistency with the third law.
\end{abstract}


\maketitle
\section{Introduction}
A very straightforward and simple derivation of a general form of Bell's inequalities was given by Wigner \cite{Wigner:1970fk} (described in detail in Sakurai \cite{Sakurai:1994uq}) whereby two independent observers make a series of spin alignment measurements on pairs of correlated particles.  This particular derivation of the inequality follows from the natural behavior of integers and the simple assumption that any correlations between the particles is classical.  While this assumption may appear obvious to many, Bell, himself, famously needed to clarify this with a discussion of one Dr. Bertlmann and his now-infamous socks \cite{Bell:2004kx}.  Thus when the inequalities are violated, the beret-wearing professor's socks reveal themselves to be non-classical oddities.  As such a very clear line is drawn between classical and quantum in this instance and Bertlmann's socks are said to be entangled (and not in the sense encountered by clothes in a washing machine).  The case of Bertlmann's socks hints at the triviality of Bell's theorem (though by no means does it minimize its importance).  What could it possibly mean, then, that some quantum systems violate a trivial statement?  The problem only becomes thornier when one considers that it has been argued that the laws governing entanglement may well be thermodynamic in nature, or, at the very least, possess thermodynamic corollaries \cite{Horodecki:2002rz,Oppenheim:2003zl}.  In particular, Wiesniak, Vedral, and Brukner have shown that entanglement is necessary in order for the third law of thermodynamics to be consistent with quantum theory \cite{Marcin-Wiesniak:2008fv}.  In this article we demonstrate that entanglement is necessary for the \emph{second} law of thermodynamics to be consistent with quantum theory by demonstrating that Bell's theorem is merely a well-disguised statement of that same law.  In the process we also demonstrate that both are trivial statements.

\section{Parallel measurements}
Suppose Alice and Bob share a pair of correlated particles.  They may each independently measure the spin projection of their individual particle within the pair.  Each measurement may be made along any one of three, \emph{not necessarily mutually orthogonal} axes, $\hat{\mathbf{a}}$, $\hat{\mathbf{b}}$, or $\hat{\mathbf{c}}$.  We refer to this setup as a \emph{single source pair} or simply a `source.'  Each particle belongs to some definite type, e.g. $(\hat{\mathbf{a}}-,\hat{\mathbf{b}}+,\hat{\mathbf{c}}+)$ meaning that if $\mathbf{S}\cdot\hat{\mathbf{a}}$ is measured, a minus sign is obtained with certainty; if $\mathbf{S}\cdot\hat{\mathbf{b}}$ is measured, a plus sign is obtained with certainty; and if $\mathbf{S}\cdot\hat{\mathbf{c}}$ is measured, a plus sign is obtained with certainty.  This is consistent with the notation in Sakurai \cite{Sakurai:1994uq}.

Each pair of particles is assumed to be correlated in some fashion (think Bertlmann's socks!) such that, for example, if Particle 1 is of type $(\hat{\mathbf{a}}-,\hat{\mathbf{b}}+,\hat{\mathbf{c}}+)$ then Particle 2 must be of type $(\hat{\mathbf{a}}+,\hat{\mathbf{b}}-,\hat{\mathbf{c}}-)$.  Table 1 demonstrates the eight possible pair types.
\begin{table}
\begin{tabular}{p{5.7cm} p{7cm}}
{\footnotesize \textbf{TABLE 1.} A list of all possible particle pair measurement outcomes given three axes along which to measure.  Adapted from Sakurai \cite{Sakurai:1994uq}.} & {\footnotesize \textbf{FIGURE 1.} A pictorial representation of $m$ sources supplying particle pairs to Alice and Bob.  Within either bin $A$ or $B$, particles of a particular type (e.g. Type 5 from Table 1) could have been obtained from multiple sources.} \\
& \\
\begin{tabular}{c c c}
\hline
\textrm{Type} & \textrm{Particle 1} & \textrm{Particle 2} \\
\hline\noalign{\smallskip}
1 & $(\hat{\mathbf{a}}+,\hat{\mathbf{b}}+,\hat{\mathbf{c}}+)$ & $(\hat{\mathbf{a}}-,\hat{\mathbf{b}}-,\hat{\mathbf{c}}-)$ \\
2 & $(\hat{\mathbf{a}}+,\hat{\mathbf{b}}+,\hat{\mathbf{c}}-)$ & $(\hat{\mathbf{a}}-,\hat{\mathbf{b}}-,\hat{\mathbf{c}}+)$ \\
3 & $(\hat{\mathbf{a}}+,\hat{\mathbf{b}}-,\hat{\mathbf{c}}+)$ & $(\hat{\mathbf{a}}-,\hat{\mathbf{b}}+,\hat{\mathbf{c}}-)$ \\
4 & $(\hat{\mathbf{a}}+,\hat{\mathbf{b}}-,\hat{\mathbf{c}}-)$ & $(\hat{\mathbf{a}}-,\hat{\mathbf{b}}+,\hat{\mathbf{c}}+)$ \\
5 & $(\hat{\mathbf{a}}-,\hat{\mathbf{b}}+,\hat{\mathbf{c}}+)$ & $(\hat{\mathbf{a}}+,\hat{\mathbf{b}}-,\hat{\mathbf{c}}-)$ \\
6 & $(\hat{\mathbf{a}}-,\hat{\mathbf{b}}+,\hat{\mathbf{c}}-)$ & $(\hat{\mathbf{a}}+,\hat{\mathbf{b}}-,\hat{\mathbf{c}}+)$ \\
7 & $(\hat{\mathbf{a}}-,\hat{\mathbf{b}}-,\hat{\mathbf{c}}+)$ & $(\hat{\mathbf{a}}+,\hat{\mathbf{b}}+,\hat{\mathbf{c}}-)$ \\
8 & $(\hat{\mathbf{a}}-,\hat{\mathbf{b}}-,\hat{\mathbf{c}}-)$ & $(\hat{\mathbf{a}}+,\hat{\mathbf{b}}+,\hat{\mathbf{c}}+)$ \\ \noalign{\smallskip}
\hline
\end{tabular} & 
\begin{tabular}{c}
\includegraphics[width=2.8in]{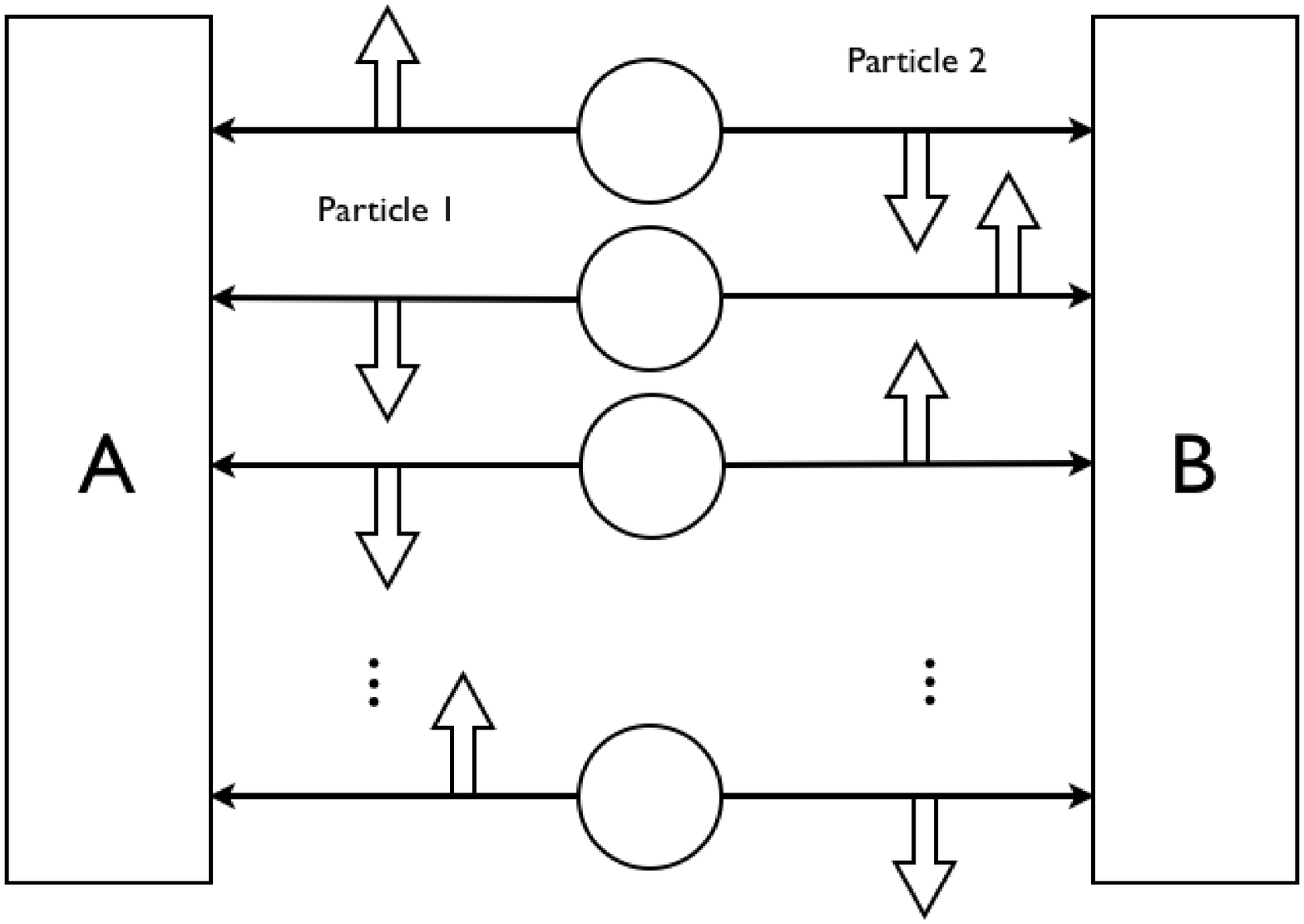}
\end{tabular} \\
\end{tabular}
\end{table}
Now assume that Alice and Bob have access to $m$ such sources allowing (but not necessarily requiring) them to make $m$ simultaneous measurements where not all of the measurements necessarily need to be the along the same axis (see Figure 1).  We also assume that the measurements are binned so that both Alice and Bob have a container consisting of a gas of particles with a mix of spin orientations.  Suppose, for example, that container $A$ contains $n$ particles of type $(\hat{\mathbf{a}}-,\hat{\mathbf{b}}+,\hat{\mathbf{c}}+)$.  These $n$ particles could all come from a single source or from multiple sources.  Thus we find that we may distinguish between different configurations, i.e. \emph{microstates}.  Each of the eight possible types of particle, then, can represent a \emph{macrostate}\footnote{It would be more accurate to call this a \emph{mesostate} since it is really an intermediary between macro- and microstates, but in the interest of preventing needless confusion we will use the more conventional \emph{macro}.}.  Note that we make a distinction between the actual physical microstates and our ability to \emph{know} them.  The \emph{multiplicity} is the number of microstates, $N_{\alpha}$, corresponding to the $\alpha$-th macrostate, i.e. the number of ways we may achieve a given macrostate.

Notice that the multiplicity is always positive definite.  As such we can form inequalities such as
\begin{equation}
N_{3}+N_{4}\le(N_{2}+N_{4})+(N_{3}+N_{7})
\end{equation}
and from them probabilities in the manner given in Sakurai \cite{Sakurai:1994uq} such as
$$
P(\hat{\mathbf{a}}+;\hat{\mathbf{b}}+)=\frac{(N_{3}+N_{4})}{\sum_{\alpha}^{8}N_{\alpha}}
$$
which gives the probability that container A's particles are spin-up along $\hat{\textbf{a}}$ while container B's particles are spin-up along $\hat{\textbf{b}}$.

Note that if each source only produces a single particle pair, the multiplicity is equal to the number of particles in a given macrostate.  Further, if we do not have access to which sources each particle in each container comes from, then this is perfectly analogous to Wigner's form of Bell's inequalities in which there is only a single source and $N_{\alpha}$ is a particle \emph{population}.  As given by Wigner we may then form a Bell inequality such as
\begin{equation}
P(\hat{\mathbf{a}}+;\hat{\mathbf{b}}+)\le P(\hat{\mathbf{a}}+;\hat{\mathbf{c}}+)+P(\hat{\mathbf{c}}+;\hat{\mathbf{b}}+).
\end{equation}
Note that we obtain (2) by multiplying (1) by the constant $1/\sum_{\alpha}^{8}N_{\alpha}$.  Thus it is really equation (1) that is fundamental here.  Since equation (1) is trivial due to the positive-definiteness of the values for $N_{\alpha}$ it would seem that Bell's inequalities are also trivial statements.  Any violation is achieved in the same manner as in Sakurai \cite{Sakurai:1994uq} since the meaning of the individual probabilities is still identical to those in Wigner's original derivation (i.e. the angles one can form between the axes still have the same interpretation), though such violations would be localized to a particular source.  In other words, one would expect, on average, that the violations would only be noticeable to Alice and Bob for small $N$.  Also note that (2) makes \emph{no distinction} between microstates.  As such, it does not include complete information of the exact state of the system but rather refers to a type of `bulk' or average behavior.  Both of these points are crucial and we will return to them in a moment.

\section{Entropy and Carath\'{e}odory's principle}
If we ignore any interactions between the particles in the containers then we can assume their orientations, as set by the measurements performed by Alice and Bob, remain unchanged.  Each particle is a dipole and thus the system in each container is analogous to a collection of ideal paramagnets.  Thus we are justified in calculating the Boltzmann entropy for a given macrostate of these particles based on their spin alignment and not some other thermodynamic quantity \cite{Schroeder:1999kx}.\footnote{Alternatively we could simply take the view that by counting microstates we are simply collecting information and thus justify ourselves by citing Jaynes \cite{Jaynes:1957fk,Jaynes:1957uq}.}  For a given macrostate composed of $i$ microstates this is
$$
S_{\alpha}=-k_{B}\sum_{i}p_{i}\textrm{ln }p_{i}
$$
where $p_{i}$ is the probability of the $i$-th microstate.  If we assume (barring the introduction of additional information) that $N_{\alpha}$ such microstates are equally probable, then the entropy of the macrostate is \cite{Schroeder:1999kx}
$$
S_{\alpha} = k_{B}\textrm{ln }N_{\alpha}.
$$
As $N_{\alpha}$, in this context, is a positive semi-definite \emph{integer}, it must also be true that ln$N_{\alpha}$ is also positive semi-definite\footnote{It is necessarily for $N$ to be an integer and not a real since the logarithm of a number between 0 and 1 is negative.}.  As such, equation (1) implies
\begin{equation}
S_{3} + S_{4} \le (S_{2} + S_{4}) + (S_{3} + S_{7}).
\end{equation}

Carath\'{e}odory's Principle, which is accepted as a valid mathematical statement of the second law of thermodynamics, states that \emph{in any neighborhood of any state there are states that cannot be reached from it by an adiabatic process} \cite{Lieb:1999vn}.  This implies that for two states, $X$ and $Y$, one can transition adiabatically from state $X$ to state $Y$ (written as $X\prec Y$) \emph{if and only if} $S(X)\le S(Y)$ in accord with Carath\'{e}odory's statement of the Second Law \cite{Lieb:1999vn}.  Thus, if we associate $S_{3}+S_{4}$ with $S(X)$ and $(S_{2} + S_{4}) + (S_{3} + S_{7})$ with $S(Y)$, then we must be able to adiabatically transition between $S_{3}+S_{4}$ and  $(S_{2} + S_{4}) + (S_{3} + S_{7})$ \emph{in accordance with the second law}.  In other words, if we assume that the latter is adiabatically accessible from the former, then (4) can be taken as a statement of the second law.

But what is adiabatic accessibility?  In classical thermodynamics the term `adiabatic' is applied to processes in which no heat is transferred\footnote{Technically it is redundant to say `heat is transferred' since, by definition, heat is any energy that is transferred across a boundary solely due to a temperature difference on either side of that boundary.}.  If we take a bin as our system and assume that the density of the gas within a bin is low enough that the particles do not interact with one another, then adding additional particles does not appreciably increase (or decrease) the temperature of the particles in the bin.  Thus it is fair to say that a transition between $S_{3}+S_{4}$ and  $(S_{2} + S_{4}) + (S_{3} + S_{7})$, which amounts to adding particles to a bin, is an adiabatic process.  Thus equation (4) is a legitimate representation of the second law.

Note that we can divide equation (3) by the partition function, $Z=\sum_{i}e^{S_{i}/k_{B}}$, to obtain equation (2).  Alternately, note that from the expansion of the terms of $Z$ we may obtain $S_{\textrm{total}}$ which we can then divide through equation (4) to obtain equation (2).  The meaning is the same as that given in the original derivation of (2) since all we have done is supply additional information in the form of the extra terms of the expansion which have then been thrown away\footnote{This interpretation is motivated by Jaynes' analysis of the maximum entropy principle \cite{Jaynes:1978vn}.}.  As such, equation (2), which we have previously identified as a form of Bell's inequalities, \emph{is essentially equivalent to a form of the second law of thermodynamics}.  Thus violations of (2) can be construed as violating the second law and we can conclude, in the manner of Wiesniak, Vedral, and Brukner \cite{Marcin-Wiesniak:2008fv}, that entanglement is a necessary feature of quantum theory that allows it to be consistent with the second law.

\section{Consequences}
Thus, given a few simple assumptions about experimental setups and the statistical distributions of particles, we have demonstrated that at least one form of Bell's inequalities is equivalent to at least one form of the second law of thermodynamics.  If all statements of the second law are equivalent to one another and all statements of Bell's inequalities are equivalent to one another then our results suggest that \emph{Bell's theorem is merely a well-disguised statement of the second law of thermodynamics}.  This is a truly profound result since it seems to suggest a route by which classicality emerges from an underlying quantum world.  As we pointed out earlier, violations only become noticeable for small $N$.  This supports the idea that the second law and Bell's theorem are emergent phenomena that describe bulk statistical behavior and are, in fact, \emph{not} fundamental.  It seems the universe really is statistical at its most fundamental level.  Nevertheless, we see that violation of the underlying inequality requires the existence of entanglement in order for there to be consistency between the quantum and classical worlds.

Finally, note that both statements are essentially trivial and yet they are violated.  This is simply further evidence that entanglement is necessary - without it the very behavior of integers might be called into question!

We thank Jan-\r{A}ke Larsson for clarifying a point about Wigner's assumptions and for pointing out that the results presented here suggested that both Bell's inequalities and the second law are actually trivial.  In addition, we thank Oleg Lunin, Ariel Caticha, John Skilling, Carolyn MacDonald, and attendees of a University at Albany (SUNY) colloquium.  In particular, we thank Kevin Knuth for his hospitality while at Albany and his invitation to speak on this topic.  This work was partially supported by a grant from the Foundational Questions Institute (FQXi).

\bibliographystyle{apsrev}
\bibliography{VaxjoPaper.bbl}

\end{document}